\documentclass[english,12pt]{article}

\usepackage{amsmath}
\usepackage{amssymb}
\usepackage{amsthm}

\usepackage[english]{babel}

\usepackage[T2A]{fontenc}
\usepackage{graphics}
\usepackage{epsfig}
\usepackage{wrapfig}

\newtheorem{theorem}{Theorem}

\theoremstyle{remark}

\theoremstyle{definition}
\newtheorem{definition}{Definition}

\theoremstyle{definition}

\newcommand {\lone} {\widehat{\mathcal L}_1}
\newcommand {\ltwo} {\widehat{\mathcal L}_2}

     \newcommand {\beq}  {\begin{equation}}
      \newcommand {\eeq}  {\end{equation}}

\newcommand{\opunit}{\text{1}\kern-0.22em\text{l}}

\newcommand{\bsi}{{\boldsymbol i}}
\newcommand{\bsj}{{\boldsymbol j}}

\newcommand{\bsv}{{\boldsymbol v}}

\author{O.German\thanks{Moscow State University, Russia. This research was  supported by
                        RFBR (grant $\textup N^\circ$ 06--01--00518) and
                        grant of the President of Russian Federation
                        $\textup N^\circ$ MK--4466.2008.1. }
\and E.Lakshtanov\thanks{Department of Mathematics, Aveiro
University, Aveiro 3810, Portugal. This research was supported by
{\it Centre for Research on Optimization and Control} (CEOC) from
the ''{\it Funda\c{c}\~{a}o para a Ci\^{e}ncia e a Tecnologia}''
(FCT), cofinanced by the European Community Fund FEDER/POCTI, and
by FCT research projects PTDC/MAT/72840/2006,
PTDC/MAT/103197/2008.}}

\title{``Minesweeper'' and spectrum of discrete Laplacians}
\date{}

\begin{document}

\maketitle

\begin{abstract}
The paper is devoted to a problem inspired by the ``Minesweeper''
computer game. It is shown that certain configurations of open
cells guarantee the existence and the uniqueness of solution.
Mathematically the problem is reduced to some spectral properties
of discrete differential operators. It is shown how the uniqueness
can be used to create a new game which preserves the spirit of
``Minesweeper'' but does not require a computer.
\end{abstract}

\section{Paper Minesweeper: history}

There is a certain class of mathematical problems which, being quite difficult to solve for an
adult mathematician in their most general setting, can be understood and even be approached to in
some particular cases by little kids. This paper is devoted to a problem of such a kind. Everybody
knows the ``Minesweeper'' computer game. A subset of a rectangular table is filled with ``mines''
and in every spare cell the number of neighbouring mines is indicated. The general problem can be
formulated as follows: given a subset of the spare part of the table with the correspondent numbers
of neighbouring mines, is there a unique way to reconstruct the original distribution of mines?

This problem in some simple cases can be used very fruitfully when teaching mathematics in primary
school, since it allows to do it while playing and does not actually require a computer. The first
experience of this kind belongs to the second author, who proposed for his pupils to reconstruct
the distribution of mines in tables of the following type:
\[ \begin{tabular}{|c|c|c|}
     \hline &2&\\ \hline &&\\ \hline 2&2&1\\ \hline
   \end{tabular} \quad \quad
   \begin{tabular}{|c|c|c|}
     \hline &2&\\ \hline &&1\\ \hline 2 & &\\ \hline
   \end{tabular} \quad \quad
   \begin{tabular}{|c|c|c|}
     \hline &&\\ \hline 1&1&1\\ \hline
   \end{tabular} \]
The result was very successful, our colleagues in several Aveiro
schools started using such tasks.

The only practical problem we had was creating new tables so that the distribution of mines would
be determined uniquely by the open area (this simplifies checking whether the solution is correct)
and that the solution would not be too easy. Eventually, we found a form of the set to be open that
guarantees the existence of a unique solution, finding which in most cases requires some thought.
As an open set we propose to take the set of staggered cells of the initial table, in the form of a
chess table. In what follows we state and prove the corresponding theorem.

    \section{Formal description of the game.}
    In the first version of the present paper \cite{first} we
    restricted ourselves to the case of rectangular fields as it is in the
    classical computer ``Minesweeper'' game. Now we decide first
    to give a formal description of ``Paper Minesweeper''. The
    reason for such a formalization is that, as our experience
    with different types of fields shows, the spirit of the game is not strictly
    connected with the rectangularity of the field. Particularly,
    our experience with tables based on the triangle tiling of the plane
    shows that the paper version of this game encounters situations
    typical for the computer ``Minesweeper'' game.

       \subsection{General case}

Let $G$ be a finite undirected graph\footnote{For the game it is
supposed     that $V$ has a certain graphical representation, such
that each     vertex represents a cell in which either a ``mine''
or a number can potentially be located.}, with vertex set $V(G)$
and edge set $E(G)$. We say that two vertices $u,v \in V(G)$ are
\emph{neighbors}, if they are connected by an edge, that is, if
$uv \in E(G)$. Thus, for every $v \in V(G)$ we can define its
\emph{neighborhood}, $N_G(v) $, as the the set of neighbors
$$
N_G(v) = \{w \in V :  vw \in E(G)\}.
$$
A pair $(A,f)$, where $A \subset V(G)$ and $f : A \rightarrow
\{0\} \cup \mathbb N$, is called an \emph{opening} if there is a
subset $M \subset V(G) \setminus A$, such that for every $v \in
A$,
\[
f(v)= |N_G(v) \cap M|, \quad v \in A.
\]

By \emph{solving} an opening $(A,f)$ we shall mean finding the
corresponding set $M$. Respectively, $A$ will be called the
\emph{set of open cells} and $M$ will be called the \emph{set of
mines}.

\begin{definition} \label{d:table}
An opening $(A,f)$ is called a \emph{table for ``Paper Minesweeper''} if and only if it admits a
unique solution.
\end{definition}

Uniqueness of the solution gives the possibility to compare the obtained answer with the right one.
Besides that, it makes the game deterministic, i.e. the presence or the absence of a mine in each
cell is predetermined.

To prepare a table one can use the following algorithm: For each
two subsets $A \subset V(G)$, $M \subseteq V(G) \setminus  A$ we
can define a function
\begin{equation}\label{fam}
f_{(A,M)}(v) = |N_G(v) \cap M|, \quad v \in A.
\end{equation}
By construction, the set $M$ solves the opening $(A,f_{(A,M)})$,
so it remains in this case to find out when the solution is
unique.

\subsection{Classical computer ``Minesweeper''}
  In the computer ``Minesweeper'' the table is an $m\times n$ rectangular subset $R$ of
  $\mathbb Z^2$, $m \leq n$:
  \[ R=\{(i,j)\ |\ 1 \leq i\leq m, 1 \leq j \leq n \}. \]
  For each $\bsi\in R$ the set of its neighbours is defined as
  \[ V_\bsi = \{ \bsi+\bsv\ |\ \bsv \in V\} \cap R, \]
  where
  \[ V=\{ (-1,1),(0,1),(1,1),(1,0),(1,-1),(0,-1),(-1,-1),(-1,0)\}. \]

\subsection{Non-formal description}
To play the game we describe in its general setting you need a
playing field (for instance, printed on paper), a writing device
(for instance, a pen) and a solution table. The playing field
consists of ``open'' cells with numbers in them and ``closed''
cells which are to be filled by a player either with symbols of
mines (crosses, for instance), or by symbols of absence of mines
(dashes, for instance). The aim of the game is to fill ALL the
``closed'' cells in such a way that each number is equal to the
number of mines in the neighbouring cells. Each playing field
should be supplemented by the definition describing which cells
are to be called ``neighbours''. According to Definition
\ref{d:table}, a table with some of the cells filled by numbers is
called \emph{a playing field for ``Paper Minesweeper''} if the
distribution of mines can be restored uniquely. This particularly
means that a player can check whether the obtained solution is
correct by comparing it with the solution table.

\section{Statement of the main theorem}

 The first result in this direction appeared within the framework
of a project of the second author's department after proposing school students of the $8$-th form
to reconstruct the distribution of mines in a table $2\times n$ with the upper string as
$A$.\footnote{Special thanks to Prof. Ana Breda for organization of this project} At that time the
following theorem was proved:

\begin{theorem} \label{t:2_times_n}
Suppose that
\[ R=\{(i,j)\ |\ 1 \leq i\leq 2, 1 \leq j \leq n \} \]
and that $n+1$ is not divisible by $3$. let $A$ be one of the two strings of $R$:
\[ A=\{ (1,i)\in R\ |\ i=1,\ldots,n \}. \]
Then for every $M\subset R\backslash A$ the opening
\[ (A,f_{A,M}) \]
admits only one solution.
\end{theorem}

To prove Theorem \ref{t:2_times_n} we used inductive calculation of $3$-diagonal determinants. The
argument is quite simple, so we skip it. Later on we obtained a more interesting theorem, which
makes the main result of this paper:

\begin{theorem} \label{t:chess}
Suppose that
\[ R=\{(i,j)\ |\ 1 \leq i\leq m, 1 \leq j \leq n \},\quad m\leq n, \]
the numbers $n+1$ and $m+1$ are coprime. Let $A$ be the subset of
$R$ in the form of a chess table:
\[ A=\{ (i,j) \in R\ |\ i+j \text{ is even}\}. \]
Then for every $M\subset R\backslash A$ the opening
\[ (A,f_{A,M}) \]
admits only one solution.
\end{theorem}

\subsection{Tables based on the triangle tiling of the plane.}

Let the neighbours of $\bsi=(i,j) \in R$ be defined by the following rule:
\[ V_\bsi = \{\bsi+\bsv\ |\ \bsv\in V^{\bsi}_t\} \cap R, \]
where
\begin{equation} \label{sosedi:tr}
  \begin{split}
    V^{\bsi}_t=\{ (0,1),(1 & ,1),(1,0),(-1,0),(1,-1),(0,-1),
    (-1,-1), \\ & (-1,1),
    (0,2),(0,-2),((-1)^{i+j},2),((-1)^{i+j},-2),\}, \quad \bsi=(i,j).
  \end{split}
\end{equation}

If we associate naturally the vertices of this graph with
triangles in the triangle tiling of the plane, then this rule
means that two triangles should be called neighbours if they have
at least one common vertex (see fig. \ref{trig}).

\begin{theorem}\label{tr}
  Let  $R=\{(i,j)\ |\ 1 \leq i\leq m, 1 \leq j \leq n \}$, and let $(m+1)$ and $(n+1)$ be not
  divisible by $4$. Let neighbours be defined by the set $(\ref{sosedi:tr})$. Suppose that $A$ is a subset of $R$
  in the form of a chess table:
  \[ A=\{ (i,j) \in R\ |\ i+j \text{ even}\}. \]
  Then for each set of mines $M\subset R\backslash A$ the opening $(A,f_{A,M})$ admits only one
  solution.
\end{theorem}

It should be noticed that, due to the larger number of neighbours, games based on such tables have
a higher level of complexity. It is curious that in these tables a player regularly encounters
situations similar to those in the computer game.

\begin{wrapfigure}{l}{45mm}
 \includegraphics[width=46mm]{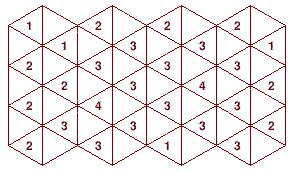}\label{trig}
 \caption{
 Table on base of equilateral triangles.}
\end{wrapfigure}

\section{An algorithm of table making.}

Before proving Theorem \ref{t:chess} we apply it to describe an algorithm of generating openings
with unique solutions. In most cases the resulting opening will be nontrivial to solve.

Suppose that $m$ and $n$ satisfying the conditions  of Theorem \ref{t:chess} are chosen. Set $M$ to
be empty initially.

1. For every $\bsi \in R\backslash A$ we execute a Bernoulli test,
and if the result is $1$, we add $\bsi$ to $M$. The probabilities
$p$ and $q$ in the test can be taken equal to $1/2$.

2. After having run through all the elements of $R\backslash A$ we define $f_{A,M}$ by \eqref{fam}
and fill all the cells $\bsi\in A$ with the values $f_{A,M}(\bsi)$.

An opening with a unique solution is ready. Enjoy the game! Here are some examples of openings that
can be obtained this way:
\begin{equation} \label{eq:3_openings}
\begin{tabular}{|c|c|c|}
\hline 1& &2\\ \hline &2&\\ \hline 1&&1\\ \hline &1&\\ \hline
\end{tabular} \quad \quad
\begin{tabular}{|c|c|c|c|c|c|}
\hline & 1& &1 & & 1 \\ \hline
 2 & &2 & &1& \\ \hline
&3& &2 & & 1 \\
\hline 1 & &3 & &2 & \\
\hline
& 1& &2 & & 1 \\
\hline
\end{tabular} \quad \quad
\begin{tabular}{|c|c|c|c|c|c|}
\hline & 1& &1 & & 1 \\ \hline
 1 & &1 & &1& \\ \hline
&1& &1 & & 1 \\
\hline 1 & &1 & &2 & \\
\hline
& 1& &1 & & 1 \\
\hline
\end{tabular}
\end{equation}
 The solutions can be found at the
end of the paper.

The first step of the algorithm can be modified in order to avoid the situation when for some $\bsi
\in A$ one has
\begin{equation}\label{easy}
\text{ either }\quad f_{(A,M)}(\bsi) =\sharp(V(\bsi)),\quad\text{ or }\quad f_{(A,M)}(\bsi)=0.
\end{equation}
We propose to fill up the first row independently and then fill all the other rows, starting with
the second one and on downwards, filling each row from left to right and taking into account the
distribution of mines already placed.

Consider the following pseudocode:
\\ for i:=2 to m
\\ for j:=1 to n
\\ if $f_{(A,M)}(i-1,j)==0$ then put a mine into the cell (i,j)
\\ if $f_{(A,M)}(i-1,j)==(\sharp(V(i-1,j))-1)$ then leave the cell (i,j) empty
\\ end
\\ end

A table obtained by the application of this algorithm can have cells satisfying the condition
\eqref{easy} only in the last row. Thus, in general, such tables are already more complicated. But
if we want the inequality $0<f_{(A,M)}(\bsi)<\sharp(V(\bsi))$ to be valid for every $\bsi \in A$ we
should apply the previous algorithm for all the rows but the last one. And for the last row we
should apply the following algorithm:
\\ for j:=2 to n
\\   \quad if $f_{(A,M)}(m-1,j)==0$ then put a mine into the cell (m,j)
\\   \quad if $f_{(A,M)}(m-1,j)==(\sharp(V(m-1,j))-1)$ then leave the cell (m,j) empty
\\  \quad if $f_{(A,M)}(m,j-1)==0$ then put a mine into the cell (m,j)
\\  \quad if $f_{(A,M)}(m,j-1)==(\sharp(V(m,j-1))-1)$ then leave the cell (m,j) empty
\\ end

    Note, however, that this procedure guarantees that there is no
    cell satisfying \eqref{easy} only in the case when the number of
    columns plus the number of rows is even. In the other case \eqref{easy} can hold for the
    cell $(m,n)$.

These algorithms were realized in \\
http://www2.mat.ua.pt/jpedro/minesweeper/the-tablep.htm


\section{Proof of Theorem \ref{t:chess}.}

Consider the opening $(A,f)$. Denote by $\mathcal X$ the set of characteristic functions
$\{0,1\}^{R \backslash A}$. We remind that there is a natural bijection between $\mathcal X$ and
the set of all the subsets of $R\backslash A$. Namely, each $M\subset R\backslash A$ corresponds to
the function $x_M \in \mathcal X$ defined by the equality
\begin{equation}
x_M(\bsi)= \left \{
\begin{array}{cc}
1, & \bsi \in M, \\
0, & \bsi \not \in M,
\end{array}
\right . \quad \bsi \in R\backslash A.
\end{equation}
Due to this bijection we can say that an element of $\mathcal X$ solves the opening meaning that
its support does. Now, the condition that a function $x \in \mathcal X$ solves the opening $(A,f)$
can be written as a system of linear equations:
\begin{equation}\label{equat}
\sum_{\bsj \in V_{\bsi} \cap (R \backslash A)} x(\bsj) = f(\bsi),
\quad \bsi \in A.
\end{equation}

Since
\[ A=\{ (i,j) \in R : i+j \text{ is even}\} \]
we can rewrite \eqref{equat} as
\begin{equation}\label{equat2}
\sum_{\bsj \in \{(i-1,j),(i+1,j),(i,j-1),(i,j+1)\} \cap R} x(\bsj)
= f(\bsi), \quad \bsi =(i,j) \in A.
\end{equation}
We fix arbitrary orders on the sets $A,R\backslash A$. Given these orders we can consider the
matrix of the system \ref{equat2} and denote it by $E$. This matrix is square, since $mn$ is even
and therefore $| A|=|R \backslash A|$. To prove the uniqueness of the solution of \eqref{equat2} it
suffices to show that $E$ is invertible. Since the invertibility of $E$ does not depend on the
order on $R$, we shall not specify the latter.

Denote
\[ P=\left  \{(x,y)=\left (\frac{\pi k}{m+1},\frac{\pi l}{n+1} \right),
   \quad (k,l) \in R \right \} \subset S^1 \times S^1. \]

Consider the real space $\mathcal L=L_2(P)$. It is well known that the system $\{ \sin kx \sin ly,
(k,l) \in R \}$ forms a basis in $\mathcal L$. Consider the operator $L$ acting on $L_2(P)$ as
multiplication by $ 2 (\cos x + \cos y)$:
\[ \frac 1 2
(Lg)(x,y)= \Big(\cos x + \cos y\Big)g(x,y), \quad g \in \mathcal L,
   \quad (x,y) \in P. \]
Note that
\[ L(\sin ix \sin jy)=\sum_{(k,l) \in \{(i-1,j),(i+1,j),(i,j-1),(i,j+1)\} \cap R} \sin kx \sin ly,
   \quad (i,j) \in R. \]
Denote
\[ \lone = {\rm span}\{ \sin kx \sin ly, (k,l) \in R, k+l \text{ is odd}\} \subset \mathcal L, \]
\[ \ltwo = {\rm span}\{ \sin kx \sin ly, (k,l) \in R, k+l \text{ is even}\} \subset \mathcal L. \]
It is easy to see that
\[ L(\ltwo) \subset \lone, \quad L(\lone) \subset \ltwo. \]
Denote
\[ L_{12}=L|_{\ltwo}, \quad L_{21}=L|_{\lone}. \]
The matrix of the operator $L_{21}$ coincides exactly with the transpose of $E$ (it is supposed
that bases in $\lone$ and $\ltwo$ are chosen in accordance with the orders on $A$ and $R\backslash
A$, which were used to define the matrix $E$). Thus it suffices to show that $L_{21}$ is
invertible.

The subspace $\lone$ is invariant under the action of $L^2$. The restriction of $L^2$ to $\lone$
coincides with $L_{12}L_{21}$. Thus, if we prove that $L^2$ is invertible, so will be both $L_{21}$
and $L_{12}$.

The spectrum of $L$ coincides with the set
\[ \Big\{2 \Big( \cos x + \cos y\Big)\ \Big|\ (x,y) \in P \Big\}. \]
To show this it suffices to notice that the functions defined by the equalities
\[ \chi_{x,y}(u,v)=\delta_{xu}\delta_{yv},\quad (x,y)\in P, \]
are eigenfunctions of $L$:
\[ \frac 1 2 L \chi_{x,y}= \Big(\cos x + \cos y\Big) \chi_{x,y}. \]
We get that $0$ is an eigenvalue of $L$ if and only if there are integers $k,l$, such that
\[ \frac{\pi k}{m+1}-\frac{\pi l}{n+1} = \pm \pi, \quad 1 \leq k \leq m, \quad 1 \leq l \leq n. \]
But this is not so, since $m+1$ and $n+1$ are coprime.

Thus, $L$ is invertible, which proves the theorem.

\section{Proof of Theorem \ref{tr}}

The proof is quite similar in this case, save that $L$ now acts as
\begin{equation}
  L(\sin ix \sin jy)=\sum \sin (i+k)x \sin (l+j)y,\quad (i,j) \in R,
\end{equation}
where the summation is taken over all the pairs
\[ (k,l)\in\{(0,1),(1,0),(-1,0),(0,-1),((-1)^{i+j},2),((-1)^{i+j},-2)\}\cap R. \]
This operator is not symmetrical and thus cannot be represented as multiplication by a function in
$L_2(P)$. Same as we did before, we can define spaces $\lone$ and $\ltwo$, and operators $L_{12}$
and $L_{21}$. Now our objective is also to find out whether $L_{21}$ has maximal rank. Indeed, by
our definition of sets $A$ and \eqref{sosedi:tr} we have $|R\backslash A|=|A|-1$ in case of odd
$mn$ and $|R\backslash A|=|A|$ in case of even $mn$ (see pict). In the latter case maximality of
rank means invertibility of $L_{21}$.

By straightforward calculations we obtain that $LL^*$ acts as
\begin{equation}
  LL^*(\sin ix \sin jy)=
  \sum_{\{(k,l) \in V_{LL^*} \cap R} \sin (i+k)x \sin (l+j)y,
  \quad (i,j) \in R,
\end{equation}
where
\[ \begin{split}
     \left(\frac{1}{2}LL^*f\right)(x,y) & =
     [3+3\cos(2x)+\cos(4x)+\cos(2y)+ \\ &
     +6\cos(x)\cos(y)+2\cos(2x)\cos(2y)+ \\ &
     +2\cos(3x)\cos(y)] f(x,y), \qquad\qquad\qquad\quad (x,y) \in P \vphantom{\bigg|}
   \end{split} \]
or
\[ \frac{1}{4}(LL^*f)(x,y) = \left | \cos(x)+\cos(y)+ e^{ix}\cos(2 y) \right |^2 f(x,y), \quad
   (x,y) \in P. \]
We have
\begin{equation}\label{eqq}
\left \{
\begin{array}{l}
\cos (x) + \cos (y) + \cos (x) \cos(2y)=0 \\
\sin (x) \cos(2 y)=0
\end{array}
\right.
\end{equation}
We cannot have $\sin(x)=0$, since $(x,y) \in P$, and so, $x=\pi k / {m+1},\ k=1,\ldots,m$. Hence
$\cos(2y)=0$, which means that $n+1$ is a multiple of $4$, since
\[ 2\frac{\pi l}{n+1} = \frac{\pi}{2}, \quad l=1,\ldots,n. \]
Thus, the second equation of \eqref{eqq} is equivalent to $\cos 2y=0$. This, together with the
first equation of \eqref{eqq}, implies that
\[ |\cos x|=\frac{\sqrt{2}}{2}, \quad (x,y) \in P. \]
Hence $m+1$ is a multiple of $4.$ Theorem \ref{tr} is proved.

\section{Solutions of \eqref{eq:3_openings}.}

\[
\begin{tabular}{|c|c|c|}
\hline 1&* &2\\ \hline -&2&*\\ \hline 1&-&1\\ \hline *&1&-\\
\hline
\end{tabular} \quad \quad
\begin{tabular}{|c|c|c|c|c|c|}
\hline -& 1& -& 1 &  *&1  \\ \hline
 2&* & 2 &- &1 &- \\ \hline
* &3 & *&2  &-  &1 \\ \hline
1 & - &3 &* & 2&*  \\ \hline - &1 &* &2 &- &1 \\ \hline
\end{tabular} \quad \quad
\begin{tabular}{|c|c|c|c|c|c|}
\hline - &1 &- & 1 &  *&1  \\ \hline
 1&* & 1 &- &1 &- \\ \hline
 -&1 &- & 1 & - &1 \\ \hline
 1& - &1 &* &2 &*  \\ \hline
 *& 1&- & 1&- & 1\\ \hline
\end{tabular} \quad \quad
\]
$
\begin{wrapfigure}{l}{45mm}
 \includegraphics[width=46mm]{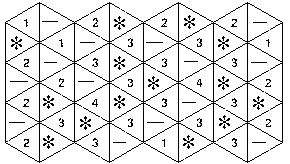}\label{trig}
 \caption{Solution for the table presented on Figure 1.}
\end{wrapfigure}
$
{\bf Acknowledgments} This research was done in frames of the EECM
 project of the mathematical department of the university of
Aveiro. This work was done during the visit of Oleg German to
Aveiro, Portugal. He thanks the CEOC research unity for warm
hospitality.

After the release of the first version of the paper E.L. had most
fruitful discussions with Alexander Klimov, Dmitry Yarotsky and
Vladimir Viro. He is also grateful to Prof. Laszlo Erd\"os for the
opportunity of giving a talk on this subject at his seminar.

\end{document}